\documentclass[pra,twocolumn, aps, showpacs, amsmath, amssymb,eqsecnum,superscriptaddress]{revtex4}
\usepackage{amsmath}
\usepackage{graphicx}
\def\myvx{{\bf x}}
\def\myvk{{\bf k}}
\def\myvp{{\bf p}}
\def\myvq{{\bf q}}
\def\A{{\bf{A}}}
\def\D{{\bf{\Delta}}}
\def\n{{\hat{\bf n}}}
\def\const{\alpha}
\def\jaf{J_{\rm AF}}
\def\jh{J_{\rm H}}
\def\dbar{{\mskip1.6mu\mathchar'26\mkern-10.6mud}}
\def\pagger{{\phantom\dagger}}
\begin{document}

\title{Double-Exchange Model:
Phase Diagram at Zero-Temperature }
\author{David Pekker}
\affiliation{Department of Physics, University of Illinois, \\ 
1110 West Green
Street, Urbana, Illinois 61801-3080, U.S.A.}
\author{Swagatam Mukhopadhyay}
\affiliation{Department of Physics, University of Illinois, \\ 
1110 West Green
Street, Urbana, Illinois 61801-3080, U.S.A.}
\author{Nandini Trivedi}
\affiliation{Department of Physics, The Ohio State University,
191 West Woodruff Avenue, Columbus, Ohio 43210, U.S.A.}
\author{Paul M. Goldbart} 
\affiliation{Department of Physics, University of Illinois, \\ 
1110 West Green
Street, Urbana, Illinois 61801-3080, U.S.A.}
\begin{abstract}
The analytical zero-temperature phase diagram of the double
exchange model for classical background spins as a function of the
carrier density and Hund's coupling in the entire range of these
parameters is presented. By constructing a continuum field theory
we explore the possibility of a continuous phase transition from
ferromagnetic state to a gently varying textured state. We find
such a transition in and below two dimensions and show that the
emerging stable state is a spin-spiral which survives the tendency
towards phase separation into commonly considered phases, and is
also energetically favored to the canted state, for low carrier
density.
\end{abstract}
\maketitle

\section{Introduction}
The Double-Exchange Model (DEM) describes the motion of noninteracting
itinerant electrons through a lattice of classical spins to which the
electron spins are coupled.  This model is relevant, e.g., in the area 
of dilute magnetic semiconductors, such a {\it GaAs\/} doped with 
{\it Mn\/}, which are important for spintronics applications~\cite{ohno}.  
In these materials, the $S=5/2$ local moments on the {\it Mn\/} sites are 
exchange-coupled to the hole carriers generated by substituting {\it Mn\/} 
for {\it Ga\/}.  While a more realistic model would include the effects of
disorder~\cite{bhatt,pinaki_disorder} arising from the random
substitution of the {\it Mn\/} atoms, as well as spin-orbit coupling
effects~\cite{macdonald}, a careful study of the DEM is a necessary
first step.  Another class of materials for which the DEM is relevant 
are the manganites, which show colossal magnetoresistance.  Although
electron-phonon interactions play an important role in the manganites
through a Jahn-Teller effect~\cite{TVR03}, some of the key features of
the strongly coupled spin, charge and lattice degrees of freedom are
captured by a DEM with lattice-distortion effects.
  
Thus, the DEM is a paradigm for a wide class of materials that show 
a strong coupling between the charge and spin degrees of freedom.
Pioneering work on this model focused on the ferromagnetic and canted 
antiferromagnetic phases at nonzero temperature~\cite{pastwork}.
Over the past few years, the DEM has received renewed attention,
stimulated in part by the numerical studies reported in
Refs.~{\cite{DagottoPRL, DagottoPRB}, where the zero-temperature
phase diagram of the simple DEM has been explored.
More elaborate extensions of the DEM---intended to move it closer
to real systems by augmenting it with physical processes such as
the super-exchange and Coulomb interactions, disorder, and
Jahn-Teller distortions, etc.---have also been investigated over
the past few years.
For example, the stability of a spin spiral state in one extension
of the model was examined in Refs.~\cite{Inoue95} and was
contrasted with the stability of the canted state.
The ground state of a model augmented with Coulomb interactions
and large Hund coupling was addressed in Ref.~\cite{Zou97}.
The instability of the homogeneous canted state with respect to
phase separation, for large Hund coupling in a model that includes
tunable super-exchange interaction, was studied in
Ref.~\cite{Kagan99}.
The phase diagram of a three-dimensional model in the infinitely
large Hund-coupling limit was studied in Ref.~\cite{Sheng98}, and
near the Curie temperature an instability with respect to
spontaneous translational symmetry breaking was proposed, as was
the possibility of phase separation.
Recently, Ref.~{\cite{Yin03}} addressed the stability of the spin
spiral state as the ground state of the DEM in the large-$S$ limit.

In spite of the considerable amount of theoretical and numerical
work on the DEM, the continuous phase transition from the
ferromagnetic phase to a spin-textured phase remains incompletely
understood.  Considerable attention has been paid to the model in
the infinite Hund-coupling regime, motivated by the fact that in
the manganites, the Hund coupling is large, compared with the
electronic bandwidth. 
The zero-temperature phase diagram has been recently studied in
Ref.~\cite{Millis01} within dynamical mean field theory (DMFT) for the
entire range of carrier concentrations and Hund coupling. It was shown
that the stable ground state in different regions of parameter space
is either a ferromagnet, or a commensurate antiferromagnet, or some
incommensurate phase with an intermediate wavevector.  Moreover, a
second-order phase transition (from the ferromagnetic to the
incommensurate phase) and a first-order transition (from the
antiferromagnetic phase to a region of phase separation) were
identified.

In this Paper, we address the magnetic ordering of the single-band
DEM with classical background spins at zero temperature (ignoring
orbital or charge ordering).  In contrast to the work of
Chattopadhyay \emph{et al.}~\cite{Millis01}, we use a continuum
field theory and a gradient expansion to determine the critical
line of continuous phase transitions separating the ferromagnetic
and textured phases in the parameter space of electron density and
Hund coupling.  This approach allows us to consider all possible
\emph{long wavelength textures} that can emerge from the DEM.  We
argue that the spin spiral is the energetically favored one from
amongst this class of textures.  As is well known, the ground
state of the DEM is phase separated at large values of the Hund
coupling.  However, by explicitly comparing the energetics of
phase separation for commonly considered textures, we argue that
there is a region of phase-space in which the continuous
transition survives the tendency towards phase separation.  We
also show that the spin spiral state is favored, energetically,
over the canted state, thereby making precise the nature of the
emergent `incommensurate state' found in phase diagrams presented
in earlier work~\cite{Millis01,DagottoPRL}.  We present the phase
diagram of the model, after including the commonly considered
candidate phases for a pre-emptive phase separation at high Hund
coupling.  In systems for which the Hund coupling is small (e.g.,
the cobaltates, diluted magnetic semiconductors, etc.) and the
disorder is low, the approach used here may be of relevance.

The present paper is organized as follows.
In Sec.~\ref{sec:AS} a continuum version of the DEM is derived.
The symmetries of this continuum version are considered and the
electronic degrees of freedom are integrated out, yielding an
effective Hamiltonian for the background spins within a gradient
expansion.
A line of continuous phase transitions is determined, and it is
shown that the spin-spiral state is the emergent stable state.
In Sec.~\ref{PhaseSeparation} the issue of whether the phase transition 
from the ferromagnetic states to the spin spiral state is pre-empted by 
another transition (to either an antiferro/ferrmagnetic 
microphase-separated state or to a canted ferromagnetic state) 
is considered.
It is shown that there is a region of the phase diagram in which
the spin spiral state is a stable state.

\section {Double-exchange model; analytical strategy}
\label{sec:AS}
The Double-Exchange Model describes noninteracting
itinerant electrons moving on a lattice of static \lq\lq
background\rq\rq\ spins whose moments are typically large compared
to that of the electron spins, and hence may be treated
classically.  The Hamiltonian is given by
\begin{eqnarray}
&&H = - t\sum_{\langle i, j \rangle \, a} \left( c_{ia}^{\dagger}
c_{ja}^{\pagger} + {\rm h.c.} \right) \nonumber \\ &&
-\frac{\jh}{S}\sum_{j \, ab} {\bf S}_j \cdot c_{ja}^{\dagger}\,
{\boldsymbol{\sigma}}_{ab}\, c_{jb}^{\pagger} +\frac{\jaf}{S^2}
\sum_{\langle i, j \rangle} {\bf S}_i\cdot {\bf S}_j ,
\label{originalH}
\end{eqnarray}
where the first, second and third terms respectively describe
electronic hopping, double-exchange and super-exchange couplings.
Here, ${\bf S}_i$ is the background spin at lattice site $i$,
which we approximate as a classical vector, $\jh/S$ is the
strength of the ferromagnetic Hund coupling, and $\jaf/S^2$ is the
strength of the antiferromagnetic coupling between the localized
spins. We shall explore the $\jh$ vs.~electron density
phase-diagram, considering $\jaf$ to be vanishingly small.
However, for technical reasons to be explained later, we shall
need $\jaf$ to be nonzero (see Appendix~\ref{app:addingAFM}).
Earlier work has mostly focused on the $\jaf$ vs.~electron density
phase-diagram of the model in the regime in which $\jh$ is very
large, so that the electrons are aligned with the background
spins.

It is convenient to transform to a new spin basis at each lattice
site, so that the local part of the Hamiltonian (i.e.~the part
that dominates at large $\jh$) is diagonal.  To this end, we
rewrite Eq.~(\ref{originalH}) in terms of the new operators $\{
d_i^{\dagger}, d_i^{\pagger} \}$ and $\{ e_i^{\dagger},
e_i^{\pagger} \}$ which are, respectively, the creation and
annihilation operators associated with the spin basis aligned and
anti-aligned with the background spin direction at site $j$:
\begin{equation}
\left( \begin{array}{c} c_{\uparrow j} \\ c_{\downarrow j}
\end{array}\right) = \left(
{\boldsymbol{\gamma}_{j}} \, \, {\boldsymbol{\gamma}_{j}^{\perp}}
\right) \left( \begin{array}{c} d_{j} \\ e_{j},
\end{array}\right),
\end{equation}
where the spinors ${\boldsymbol{\gamma}_j}$ and
${\boldsymbol{\gamma}_j^{\perp}}$ are defined in terms of the the
polar and azimuthal angles of the background spin direction at
site $j$, i.e., $\theta_i$ and $\phi_i$, and are given by
\begin{subequations}
\begin{eqnarray}
{\boldsymbol{\gamma}_j} &=& e^{+i \chi_j} \left(
\begin{array}{c} \phantom{-}e^{-i \phi_j /2} \cos \frac{\theta_j}{2} \\ \phantom{-}e^{+i \phi_j/2} \sin
\frac{\theta_j}{2} \end{array} \right) \\
{\boldsymbol{\gamma}_j^{\perp}}  &=& e^{-i \chi_j} \left(
\begin{array}{c} - e^{-i \phi_j/2} \sin \frac{\theta_j}{2} \\
\phantom{-}e^{+i \phi_j/2} \cos \frac{\theta_j}{2}
\end{array} \right).
\end{eqnarray}
\end{subequations}
${\boldsymbol{\gamma}}$ and ${\boldsymbol{\gamma}}_{\perp}$ form
an orthonormal local basis corresponding to the aligned and
anti-aligned spin states.  This mapping is defined up to a phase
factor $\chi_j$, which is a gauge freedom~\cite{assa}. In the
continuum limit, the Hamiltonian is (up to the corresponding gauge
transformation) given by
   \begin{widetext}
\begin{eqnarray}
\label{ContinuumH} H = \!\!&-& \!\!\!\! \int \!\! d\myvx \Bigg[
\psi^{\dagger}(\myvx) \left\{\left(\partial_{\alpha} - iA_{\alpha}
\right)^2 +
       \frac{1}{4}\left(\partial_{\beta} \n \cdot \partial_{\beta} \n \right) - \jh  \right\}\psi(\myvx)
   - \varphi^{\dagger} (\myvx)\left\{ \left(\partial_{\alpha} + iA_{\alpha} \right)^2
       \varphi(\myvx) + \frac{1}{4}\left(\partial_{\beta}\n \cdot \partial_{\beta} \n \right) + \jh
       \right\}\varphi(\myvx) \nonumber \\
   &&\qquad\qquad+\Big\{ \psi^{\dagger}(\myvx) \,\,
\Delta_\alpha(\myvx)\,\,
\partial_\alpha \varphi(\myvx)
\, \,  + \, \frac{1}{2} \psi^{\dagger}(\myvx) \left(
\partial_\alpha \Delta_\alpha(\myvx) \right) \varphi(\myvx) + {\rm h.c.} \Big\} - \jaf \left(\partial_{\beta} \n \cdot \partial_{\beta} \n \right)
\Bigg] ,
\end{eqnarray}
\end{widetext}
where $\psi(\myvx)$ and $\varphi(\myvx)$ are, respectively, the
field (annihilation) operators describing locally aligned and
anti-aligned electrons and $\myvx$ is the position vector,
$\n(\myvx)$ is the unit vector along the background spin
direction, and $A$ and $\Delta$ are vector potentials originating
from a Berry phase and are defined by
\begin{subequations}
\begin{eqnarray}
\label{defA} A_\alpha(\myvx) &=& \frac{1}{2} \cos \theta(\myvx)
\,\,
\partial_\alpha \phi(\myvx) \\
\label{defD} \Delta_\alpha(\myvx) &=& -\partial_\alpha \theta
(\myvx) +i \sin \theta(\myvx) \,\,
\partial_\alpha \phi (\myvx).
\end{eqnarray}
\end{subequations}
Note that $\Delta_\alpha^* \Delta^{}_{\alpha} =
\partial_{\beta} \n \cdot \partial_{\beta} \n$.
Also note that in passing to the continuum limit we have retained
only the lowest-order terms in the gradient expansion of the field
operators.  As a result, the electron kinetic energy, and hence
the band structure that corresponds to it, is of a simple
parabolic form.

\subsection{Symmetries of the continuum Hamiltonian}

\subsubsection{Local gauge invariance}

The mapping from a vector (representing a classical spin) in
three-dimensional space to a spinor in $SU(2)$ is defined up to an
angle $\chi(\myvx)$, an overall phase factor, which is a $U(1)$
gauge freedom~\cite{assa}, i.e., under the simultaneous
transformations
\begin{equation}
\begin{array}{lcl}
\label{gauge-transform} \A(\myvx) \rightarrow \A(\myvx) + \nabla
\chi(\myvx) & &  \D(\myvx) \rightarrow e^{2i \chi(\myvx)}
\D(\myvx) \\ \noalign{\medskip} \psi(\myvx) \rightarrow e^{i
\chi(\myvx)}\psi(\myvx) & & \phi(\myvx) \rightarrow e^{-i
\chi(\myvx)} \phi(\myvx)
\end{array}
\end{equation}
the Hamiltonian is invariant.

\subsubsection{Global spin rotation invariance}

The Hamiltonian is also invariant under rotation of all the
background spins by a global angle. One can show that the effect
of a global rotation is identical to a gauge transformation. For
example, let us consider rotations about $x$-axis by a small angle
$\omega$. The unit vector $\hat{n}(\myvx)$ transforms to
$\hat{n}'(\myvx)$ where $ \theta'(\myvx) \approx \theta(\myvx) -
\omega \sin \phi (\myvx)$ and $\phi'(\myvx)  \approx \phi (\myvx)
- \omega \cot\theta (\myvx) \cos \phi(\myvx)$. Notice that the
above transformation leaves $
\partial_{\beta} \hat n \cdot \partial^{\beta} \hat n $ invariant, but
the vectors $\A(\myvx)$ and $\D(\myvx)$ transform, up to first
order in $\omega$, in the following manner:
\begin{subequations}
\begin{eqnarray}
\label{rotation-gauge} A_\alpha &\rightarrow & A_\alpha -
\frac{\omega}{2}\,\partial_\alpha \left(\frac{\cos \phi}{\sin
\theta}\right),\\  \Delta_\alpha &\rightarrow &  e^{-i \omega \cos
\phi/\sin \theta} \Delta_\alpha.
\end{eqnarray}
\end{subequations}
On identifying the factor $-\omega \cos \phi/ 2 \sin \theta$
with the gauge parameter $\chi$ [see Eq.~(\ref{gauge-transform})],
we see that Eq.~(\ref{rotation-gauge}) describes a gauge
transformation corresponding to rotations. A similar analysis can
be carried out for rotations about the $y$-axis; for rotations
about the $z$-axis the transformation is trivial. Therefore,
global rotational-invariance corresponds to the gauge freedom in
the model.

\subsection{Effective Hamiltonian}
\label{sec:effham}

In this section we derive an effective Hamiltonian $H_\text{eff}$
governing the spatially-dependent background spin orientation
(\emph{texture}), in the limit that the texture varies on
length-scales much bigger than the inverse Fermi wavevector. To
do this, we integrate out the electronic degrees of freedom,
assuming that the zeroth-level description corresponds to a system
in the presence of a spatially uniform texture.  The contribution
due to any inhomogeneity of the texture is then treated as a
perturbation via a gradient expansion.  The demerit of this
continuum approach is that background spin configurations that
change abruptly from one site to another (e.g.~canted states or
antiferromagnetic state) are excluded from consideration.

The effective Hamiltonian that results from this approach is a
functional of $\A(\myvx)$, $\D(\myvx)$ and $\n(\myvx)$ and their
derivatives. Working at fixed chemical potential $\mu$, the
effective Hamiltonian is defined as
\begin{eqnarray}
\!\!\!\!\!\!e^{-\beta H_\text{eff}[\A, \D, \n, \mu]}\!\! &&\equiv
\!\!\!\int \! {\cal{D}}\psi {\cal{D}} \varphi \,\, e^{-\beta\left(
H[\psi, \varphi, \A, \D, \n] - \mu N \right)}\nonumber
\\ = \int \! &&\!\!\!\!{\cal{D}}\psi {\cal{D}} \varphi \,\,
e^{H_0[\psi, \varphi, \mu] + H_1[\psi, \varphi, \A, \D, \n]},
\end{eqnarray}
where $ N \equiv \int d\myvx \left( \psi^{\dagger}(\myvx)
\psi(\myvx) + \varphi^{\dagger}(\myvx) \varphi(\myvx) \right)$,
$H_0$ is the free Fermi gas Hamiltonian, and the perturbation
$H_1$ is a functional of $\A$, $\D$ and $\partial \n$, each of
which has one spatial derivative [see
Eq.~(\ref{defA},\ref{defD})], and therefore is small in the sense
of our approximation scheme. The form of $H_\text{eff}$ is
constrained by gauge invariance. Thus, keeping allowed terms to
quartic order in gradients, but for now  setting $\jaf$ to zero,
we find the following form, arranged in increasing order:
\begin{eqnarray}
\label{effective-H} && H_\text{eff}[\A(x), \n(x), \D(x)] \nonumber
\\ &&= \int d\myvx \Big\{a(\mu)\,
\partial_{\alpha}\n \cdot \partial_{\alpha}\n+ b(\mu) \left(
\partial_{\alpha} \n\cdot \partial_{\alpha}\n \right)^2 \nonumber \\
&&+ \,\, c(\mu) F_{\alpha \beta}F_{\alpha \beta}\! + d(\mu) \vert
D_\alpha \Delta_\alpha \vert^2\! + e(\mu) \vert D_\alpha
\Delta_\beta \vert^2\Big\} \nonumber \\ &&+ \cdots,
\end{eqnarray}
where the coefficients $a(\mu)$, $b(\mu)$, $c(\mu)$, $d(\mu)$ and
$e(\mu)$ are evaluated by computing the corresponding Feynman
diagrams. Not surprisingly, the results are compactly expressed in
terms of the following quantities: $F_{\alpha \beta}\equiv
\partial_{\alpha} A_{\beta}(x)-\partial_{\beta} A_{\alpha}(x)$
and $D_\alpha=\partial_\alpha - 2i A_\alpha$; note that the
combination $D_\alpha \Delta_\alpha$ is gauge invariant.

The Feynman diagrams contributing to the lowest-order terms in the
gradient expansion are shown in Fig.~\ref{lowest-order}.  The
amplitudes of these diagrams are both proportional to
$\left(\partial^{\alpha} \n \right)^2$. Therefore, their coefficients,
\(\rho_1\) and \(\rho_2\), add to give the stiffness of the
ferromagnetic state \(\rho\); when \(\rho\) goes negative, the
ferromagnetic state becomes linearly unstable. The dependence of the
corresponding limit of stability on $d$, $\mu$ and $\jh$ is among the
central results of this paper.
\begin{figure}[h]
1.\includegraphics[width=2.2cm]{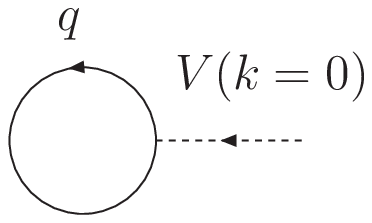} \hskip1cm
2.\includegraphics[width=2.8cm]{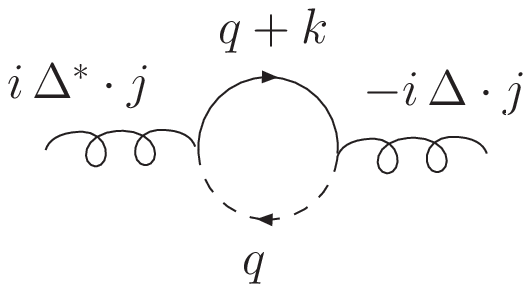}
\caption{\label{lowest-order} Lowest-order Feynman diagrams (1 and 2):
solid lines represent aligned electron propagators, dashed lines
represent anti-aligned electron propagators, curly lines
represent \(\Delta_\alpha\), and dotted lines represent
\(\partial_{\alpha} \n\cdot \partial_{\alpha}\n\).}
\end{figure}
The contribution from these two diagrams compete with one another.
The first, gives a positive contribution to the energy, because a
spatial variation of the background spins decreases the hopping
amplitude, via the Anderson-Hasegawa mechanism: $t \rightarrow t
\cos(\Theta_{ij}/2)$ where $\Theta_{ij}$ is the angle between the
nearest-neighbor spins $i$ and $j$ in the discrete version of this
model. The second diagram gives a negative contribution; spatial
variations in the background spin orientation allow for mixing of
aligned and anti-aligned bands, thereby lowering the energy.  The
contributions to the stiffness [i.e.~the coefficient of
$(\partial_{\alpha}\n)^2 $] are:
\begin{subequations}
\label{eq:smart}
\begin{eqnarray}
\rho_1&=&\frac{ (\mu+\jh)^{d/2} + (\mu-\jh)^{d/2} \, \Theta [ \mu-\jh ]
}{ 2^{1+d} \, \pi^{d/2} \, d \, \Gamma \left[ d/2 \right] },
\qquad\qquad\quad{}
\\
\rho_2&=&\frac{(\mu+\jh)^{1+d/2}-(\mu-\jh)^{1+d/2} \,
\Theta[\mu-\jh]}{2^{d+1} \pi^{d/2} (2+d) \jh \Gamma [1+d/2]},
\end{eqnarray}
\end{subequations}
where $\Gamma[\cdot]$ is the Gamma function, $\Theta [\cdot]$ is
the Heaviside step function, and $d$ is the dimension of space.

By examining the stiffness as a function of $\mu$ and $d$ (see
Fig.~\ref{fig:dimdep}) we observe that $d=2$ is a threshold
dimension, in the sense that the instability occurs for dimensions
less than two (but not for dimensions greater than two).  We
emphasize that the precise location of the transition, as well as
the threshold dimension, depends on the form of the bare
electronic band structure, which we have taken to be parabolic.
Corrections to the parabolic electronic dispersion relation would
alter both the location of the transition and the threshold
dimension.

\begin{figure}
\vskip0.3cm
\includegraphics[width=6cm]{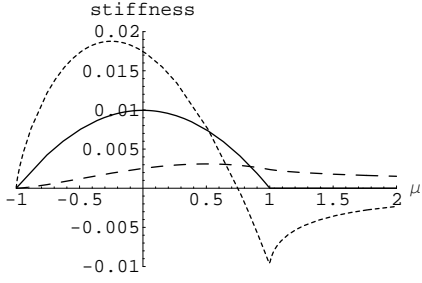}
\caption{Ferromagnetic stiffness in dimension $d$ for $d=1.5$
(dotted), $d=2$ (solid), and $d=3$ (dashed). Here, \(\jh=1\) and
\(\jaf=0\). Note that for $d=2$, the stiffness becomes zero at
\(\mu=\jh\) but does not become negative.} \label{fig:dimdep}
\end{figure}
In particular, in two dimensions the contributions combine to give
the stiffness
\begin{equation}
\frac{\jh^2 - \mu^2 }{32 \pi \jh}.
\end{equation}
This shows that in two dimensions and at zero temperature there is
a critical chemical potential $\mu_{\rm c}=\jh$ above which the
ferromagnetic phase loses stability and, as we shall see,
undergoes a transition to a `textured' phase.  This critical
chemical potential coincides with the bottom of the anti-aligned
electron band. To investigate how the instability of the
ferromagnetic state is resolved, it would therefore be necessary
to raise the chemical potential above its critical value, which
would begin populating this band.  However, as the gradient
expansion is an expansion powers of
\(q_\text{texture}/k_\text{F}^\text{upper}\), it would not
converge, as $k_\text{F}^\text{upper}$ would be very small. We
approach this dilemma by noting that these results were obtained
in the absence of an antiferromagnetic term in the original
Hamiltonian~\ref{originalH}, i.e., for $\jaf=0$.  The precise
location $\mu_{\rm c}(\jh,d)$ of the instability of the
ferromagnetic state is perturbed, and in general shifted to lower
value, in the presence of a positive $\jaf$, thus creating a
region in the $(\mu,\jaf)$ plane in which the ferromagnetic state
has become unstable and yet $\mu$ is still smaller than $\jaf$,
sothat the upper band remains unoccupied. This scheme opens up a
region of the phase diagram in which our gradient expansion
remains valid and, at the same time, a textured state is prefered.

In order to investigate the form of the (stable) textured state
that replaces the (unstable) ferromagnetic state at chemical
potentials immediately greater than the critical one (given in two
dimensions by $\mu=\jh$), we expand the effective Hamiltonian
density to fourth order in gradients of $\n$, and minimize it with
respect to all textures that vary only on lengthscales longer than
the Fermi wavelength.  Via the extension to quartic order of the
diagrammatic expansion described in the present section, and for
the case of two dimensions, we find the effective Hamiltonian to be
\begin{eqnarray}
H_\text{eff} &=& \frac{\jh^2 - \mu^2}{32 \pi \jh}
(\partial_{\alpha} \n)^2 + \frac{\mu( \jh^2 + \mu^2) }{ 256 \pi
\jh^3} (\partial_{\alpha} \n)^4 \nonumber \\ &+&
\frac{3\jh^2(\jh+\mu) - \mu^3 }{48 \pi \jh^3} F^{\alpha \beta}
F_{\alpha \beta} \nonumber
\\ &+& \frac{(\jh + \mu)^2 (2 \jh - \mu)}{ 192 \pi
\jh^3} |D_{\alpha}\Delta_{\beta}|^2  \nonumber
\\ &-& \frac{\jh^3 - \mu^3}{96 \pi
\jh^3}|D_{\alpha}\Delta^{\alpha}|^2-\jaf (\partial_{\alpha} \n)^2,
\label{effectiveH2}
\end{eqnarray}
where the terms associated with $\jaf$ arise from the
antiferromagnetic term in Eq.~(\ref{originalH}).

The details of making this extension to quartic order are given in
Appendix~\ref{app:diagrams}.  For $\mu$ larger than its critical
value, the coefficient of the first term in $H_\text{eff}$
(e.g.~the ferromagnetic stifness) is negative, and therefore it is
favorable for the ground state to have a non-uniform texture.  We
now show that the remaining terms, which are fourth order in
gradients and serve to restabilize the textured state, are all
positive-definite whenever \(0<\mu<\jh\). The coefficients of
$(\partial_{\alpha} \n)^4$ and $F_{\alpha \beta}F_{\alpha \beta}$
are positive for $0< \mu < \jh$, ensuring that these terms are
indeed positive-definite.  If we neglect the surface terms, the
forth and fifth terms can be recast in the following form:
\begin{equation}
\frac{\mu \left( \jh^2 - \mu^2 \right)}{64 \pi \jh^3
}|D_{\alpha}\Delta_{\beta}|^2 + \frac{\jh^3 + \mu^3}{48 \pi \jh^3}
\sin^2 \theta \left\vert \nabla \theta \times \nabla \phi
\right\vert^2.
\end{equation}
In this form, the coefficients of each of these terms is positive for
$0< \mu < \jh$, and therefore all the fourth-order terms are indeed
positive-definite.  Keeping only the first two terms, it is easy to
check from the differential equation for the ground state, which
follows from varying $H_\text{eff}$ with respect to the fields
$\theta$ and $\phi$, that a spin spiral state [e.g.~$\theta=\pi/2$ and
$\phi={\bf q}\cdot{\bf x}$] (where ${\bf q}$ is a suitably-chosen
wavevector) minimizes the energy.  The third, forth and fifth terms
vanish for the spiral state.  Hence, the spiral state is a local
minimum of the energy, as small perturbations around it would
certainly raise the first two terms, and can only increase effect the
remaining terms (as they are zero to begin with). The implication of
this analysis is as follows: in the Double-Exchange Model there is a
region of the zero-temperature $(\mu,\jh)$ phase diagram in which a
spin spiral state is (at least locally) a stable ground state.  This
state emerges on the high-$\mu$ (or, equivalently, low-$\jh$) side of
the continuous phase transition line, on the other side of which the
ferromagnetic state is the stable state.  By using the fourth-order
terms to restabilize the instability caused by the negative stiffness
and including the effects of \(\jaf\), we find that the wavevector
\(\alpha\) of the spiral is given by
\begin{equation}
\const^2 = \frac{4\jh^2(\mu^2 - \jh^2 + 32 \pi \jh
\jaf)}{\mu(\mu^2 + \jh^2)}
\approx {\frac{4(\mu-\mu_\text{c})}{1-16 \pi (\jaf/\jh)}},
\end{equation}
where \(\mu_\text{c}^2 = \jh^2 (1-32 \pi \jh \jaf)\). The approximate
form holds for \(\mu\agt\mu_\text{c}\). 

Here, we note that if \(\jaf\) is greater than zero then the
spin-spiral would not persist to arbitrary small carrier density, as
the anti-ferromagnetic state becomes stable.  Is such a continuous
transition pre-empted by a first-order transition into a
microphase-separated state? We explore this possibility in the next
section.

\section{Phase separated and canted states}
\label{PhaseSeparation}

\subsection{Collinear magnetic states}

In this section we compare the energies of several
commonly-studied types of microphase-separated
states~\cite{Golosov02} in double-exchange magnetic systems, with
the aim of comparing their stability relative to the spin spiral
state.  By microphase-separated we mean states that have
meso-scale structure (magnetic and/or electronic) controlled by a
competition between long-range interactions and interfacial
energies.  Prior work~\cite{Golosov02,DagottoPRB} has focused on the
competition between the super-exchange and double-exchange
coupling strengths, and has commonly assumed the latter to be
infinite, or at least very large.  In the present setting, we are
concerned with the {\it entire\/} range of double-exchange
coupling strengths, but only with small super-exchange coupling
strengths.

In order to have the coexistence of the ferromagnetic and
antiferromagnetic microphases associated with the microphase
separation that we are considering, their thermodynamic and
chemical potentials should coincide with one
another~{\cite{Nagaevbook}}.  These condition are necessary for
microphase coexistence, but they are not sufficient in settings
involving long-range interactions, such as those due to distinct
charge-densities in the coexisting microphases, or interface
energies associated with regions separating microphases. However,
by examining the complement of the regions of the phase diagram
that satisfy the aforementioned necessary conditions or are
antiferromagnetic, we can locate the regions in which the
homoegenous ferromagnetic state or the textured state have a
chance of being stable.  (As we shall be limiting our
consideration to the various types of antiferromagnetic ordering
listed in Fig.~{\ref{block}}, we may fail to exclude some regions
of the phase diagram that we shall be calling ferromagnetic or
textured.)

\begin{figure}
\includegraphics[width=8cm]{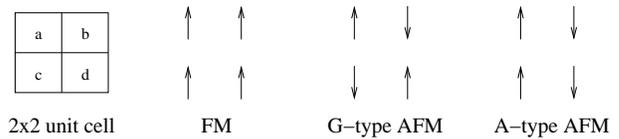}
\caption{Arrangements of background spins on the $2 \times 2$
plaquettes that are considered in the text. } \label{block}
\end{figure}
We proceed by locating those regions of the $(\jh,n)$ phase
diagram in which either the conditions for microphase-coexistance
are satisfied or there is an antiferromagnetic state of lower
energy than the ferromagnetic state.  The single-electron band
structure of the DE Hamiltonian~(\ref{originalH}) for ferromagnet
(F), G-type (G) and A-type (A) antiferromagnet are given by For
ferromagnet:
\def\remspace{\!\!\!\!\!\!\!\!\!\!\!\!\!\!\!}
\begin{subequations}
\begin{eqnarray}
\remspace\text{(F)}\quad \epsilon_k&=&\pm \jh \pm 2 t \cos k_x \pm
2 t \cos k_y - \mu,
\\
\remspace\text{(G)}\quad \epsilon_k&=&\pm \sqrt{\jh^2 + 4 t^2
(\cos k_x \pm \cos k_y)^2} - \mu,
\\
\remspace\text{(A)}\quad \epsilon_k&=&\pm 2t \cos k_x \pm \sqrt{
\jh^2 + 4 t^2 \cos^2 k_y} - \mu.
\end{eqnarray}
\end{subequations}
For both antiferromagnetic arrangements (G and A), all the
eigenvalues are doubly degenerate.  The energy density and the
electron density at zero temperature are given by
\begin{subequations}
\begin{eqnarray}
E&=& \int\limits_{\text{BZ}}\frac{d^{2}k}{4 \pi^{2}}\,
\Theta(\epsilon_k)\,\epsilon_k ,\\ N&=&
\int\limits_{\text{BZ}}\frac{d^{2}k}{4 \pi^{2}}\,
\Theta(\epsilon_k),
\end{eqnarray}
\end{subequations}
where for the antiferromagentic cases the Brillouin zone should be
halved (in each direction), relative to the ferrmoagnetic case.

From these ingredients we find numerically the lines in the
\((\mu,(\jh))\) plane at which the A-AFM/FM and G-AFM/FM phase
transitions occur.  On these phase boundaries, for each pair of
competing states we calculate a pair of lines $n_{\rm c}(\jh)$
corresponding to the density of each state.  On the $(\jh,n)$
phase diagram these lines bound the regions depicted in
Fig.~\ref{phasediagram} inside which the two competing microphases
have a chance of coexisting.

\begin{figure}
\vskip0.5cm
\includegraphics[width=8cm, angle=0]{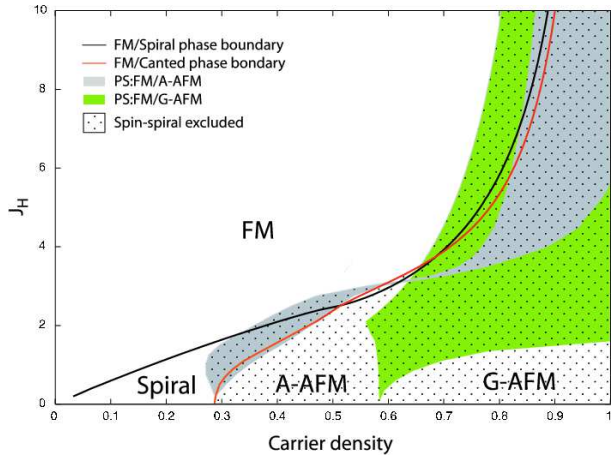}
\caption{Dependence of the ground state of the Double Exchange Model
on the carrier density and the Hund coupling at zero antiferromagnetic
coupling. At low carrier densities and high Hund couplings
ferromagnetism is favored. In certain other regions, distinct
homogeneous states are favored, including spin spiral (see
Fig~\ref{fig:spinspiral}), A- and G-type antiferromagnetism (see
Fig.~\ref{block}). In yet other regions several types of FM/AFM
microphase-separated states are favored. As explained in the text,
the boundaries indicated for microphase-separated states are in fact
stability limits; the true boundaries must lie within these stability
limits. Only in undotted region can one be certain that the
spin-spiral state is the favored state. (Color version available
online.) }
\label{phasediagram}
\end{figure}

\subsection{Uniform canted magnetic states}

\begin{figure}
\vskip0.5cm
\includegraphics[width=2cm, angle=0]{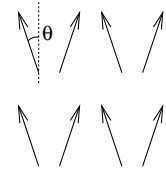}
\caption{Background spin configuration in the canted ferromagnetic
state.} \label{fig:cantedFM}
\end{figure}
Next we consider the instability of the ferromagnetic state with
respect to the canted ferromagnet state, in order to determine the
phase boundary between them.  Repeating the proceedure of the
previous subsection, we find the single-electron band structure of
Eq.~(\ref{originalH})
\begin{equation}
\begin{split}
\epsilon_k &= - 2t \cos k_y - \mu \\ &\pm \sqrt{ \jh^2 + 2t^2 ( 1
+ \cos 2 k_x) \pm 4 \jh t \cos k_x \cos \theta },
\end{split}
\end{equation}
where $\theta$ is the canting angle of the background spin with
respect to the $z-$axis, as shown in Fig.~\ref{fig:cantedFM}.  To find
the phase transition line in the \((\mu,\jh)\) diagram, we determine
when the stiffness of the ferromagnetic state with respect to canting
becomes zero.  The resulting phase-boundary line is shown in
Fig.~\ref{phasediagram}. From the diagram, it appears that the canted
state is always pre-empted by a microphase-separated state.

\section{Concluding remarks}

To study smoothly varying textures of the Double Exchange Model, we
have followed the familiar program of expanding the free energy of the
model in powers of gradients of the background spin texture and
tracing out the electronic degrees of freedom. This program has wide
applicability to the study of long wavelength patterns in lattice
models such as the Double Exchange Model. The main result of our paper
is the phase diagram~\ref{phasediagram} for the Double Exchange Model,
which we have obtained analytically as a function of the carrier
density and the Hund coupling $\jh$ between the carrier spins and the
lattice of classical backgropund spins.  Through the application of
this program, we find that the spin-spiral state is indeed a stable
state for low carrier-densities and has a continuously-varying
wavevector $\alpha\sim |\mu-\mu_\text{c}|^{1/2}$.  By direct
diagonalization we also find that the transition from the
ferromagnetic state to the canted state is essentially pre-empted by
phase separation into different types of antiferromagnetic states.

\acknowledgements We acknowledge helpful discussions with
S.~L.~Cooper, D.~I.~Golosov, R.~M.~Martin, H.~R.~Krishnamurthy,
T.~V.~Ramakrishnan, and M.~B.~Salamon. We are especially graetful to
B.~H.~Lee for sharing with us unpublished numerical work of his on the
DEM, which prompted us to consider the isues addressed in the present
paper.  This work was supported by the U.S.~Department of Energy,
Division of Materials Sciences under Award No.~DEFG02-91ER45439,
through the Frederick Seitz Materials Research Laboratory at the
University of Illinois at Urbana-Champaign.

\appendix

\section{Simple model of spin spiral texture in two dimensions}
\label{app:addingAFM}

\begin{figure}[h]
\vskip0.5cm
\includegraphics[width=5cm, angle=0]{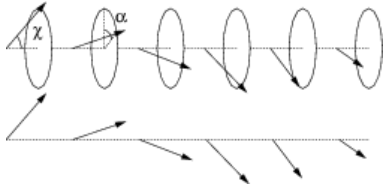}
\caption{\label{fig:spinspiral} Spin spiral configuration of
background spins, with cone angle \(\chi\) and wavevector \(\alpha\).}
\end{figure}

In this appendix, we solve for the exact ground-state energy of
the DEM for the spin spiral state as the background spin texture.
We confirm that in two dimensions the critical line for the
transition lies at $\mu = \jh$, as stated in
Sec.~\ref{sec:effham}. The integrating-out of the electronic
degrees of freedom, which we have carried out in the main text to
determine the ground-state energy, becomes more transparent
through this example, in which the process is nonperturbative.
The simplification ensues when on restricts attention to a
specific class of background spin configurations, viz., spin
spirals, which have the form
\begin{equation}
\theta(x) = \chi \,\, ; \,\, \phi(x) = \const \cdot x\,,
\end{equation}
where $\const$ is a constant wavevector.  With this choice, $\A(x)
=\const/2 \cos \chi $, $\D(x) = i \const \sin \chi $ and
$(\partial\n)^2 = \const^2 \sin^2 \chi$. The
Hamiltonian~\ref{ContinuumH} then reduces to
\begin{widetext}
\begin{eqnarray}
H_{\text{2D}} \!=\! \text{\Large $\int$} 
\begin{array}{cc}
\big(\psi^{\dagger}(k) & \varphi^{\dagger}(k)\big)\\
\noalign{\smallskip} &
\end{array}
\!\!\! \left(
\begin{array}{cc} { k^2 + \const \cdot k \cos \chi + \const^2/4 - \jh - \mu} & -\const \cdot k
\sin \chi \\ \noalign{\smallskip} -\const \cdot k \sin \chi & {k^2
- \const \cdot k \cos \chi + \const^2/4 + \jh - \mu }
\end{array} \right) \left( \begin{array}{c} \psi(k) \\
\noalign{\smallskip} \varphi(k)
\end{array}
\right) \frac{d^{2}k}{4\pi^2}.
\end{eqnarray}
\end{widetext}
On diagonalizing this Hamiltonian and calculating the effective
energy for the spin spiral state, we find that the critical
chemical potential is equal to the Hund coupling.  When $\mu$ is
larger than $\jh$ the ferromagnetic state is unstable with respect
to the formation of a spin spiral state.
For $\chi = \pi/2$ and $\mu < \jh $ the energy has the form
\begin{equation}
E_\text{eff} = - \frac{(\jh + \mu)^2}{8 \pi} + \const^2 \,\frac{
\jh^2 - \mu^2}{32 \pi \jh} + \const^4 \, \frac{\mu (\jh^2 +
\mu^2)}{256 \pi \jh^3}.
\end{equation}
For $\mu>\jh$, minimizing $E_{\text{eff}}$ with respect to
$\const$ gives
\begin{equation}
\const^2  =  \, \frac{4\jh^2(\mu^2 - \jh^2)}{\mu(\mu^2 + \jh^2)},
\end{equation}
which determines the pitch of the stabilizing spin spiral.

\begin{figure}[h]
\vskip0.5cm
\includegraphics[width=6cm, angle=0]{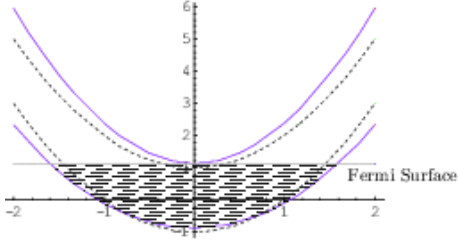}
\caption{\label{fig:DEband} Band structure in the presence (solid
line) and absence (dashed line) of the spin-spiral magnetic state in
two dimensions.  Notice that the Fermi surface just touches the bottom
of the anti-aligned band for the optimal spin-spiral state. }
\end{figure}

\begin{widetext}

\section{Evaluation of Feynman diagrams}
\label{app:diagrams}

In this appendix we present a sample diagram calculation by
considering the two diagrams of lowest order in the gradient
expansion (see Sec.~\ref{sec:effham}).  The Matsubara Green
functions for aligned ($\psi$) and anti-aligned ($\varphi$)
electrons are
\begin{gather}
G_{\psi}(\myvp,i p_n)   =\frac{1}{i p_n-(p^2-\mu-\jh)},\\
G_{\varphi}(\myvp,i p_n)=\frac{1}{i p_n-(p^2-\mu+\jh)}.
\end{gather}
The two diagrams that contribute to the ferromagnetic stiffness
are diagrams $1$ and $2$ in Fig.~\ref{fig_diag}. The amplitude
corresponding to the first diagram is
\begin{equation}
\frac{1}{\beta} \sum_{i q_n} \int \dbar\myvq \,
\left\{G_{\psi}(\myvq,i q_n)+G_{\varphi}(\myvq,i q_n)\right\}
\frac{1}{4} \left(\partial_{\beta}\n \cdot
\partial_{\beta} \n \right),
\end{equation}
where $\dbar\myvp$ stands for $d^d p/(2 \pi)^d$.  The integral
over the internal momentum results in the following expression,
which is essentially the the sum of the volumes of two
$d-$dimensional spheres of respective radii $\sqrt{\mu-\jh}$ and
$\sqrt{\mu+\jh}$:
\begin{equation}
\frac{2^{-1-d} \pi^{-d/2} \mu^{d/2}} {d \Gamma(d/2)}
\left\{\Theta(\mu+\jh)(\mu+\jh)^{d/2}
+\Theta(\mu-\jh)(\mu-\jh)^{d/2} \right\}.
\end{equation}
The amplitude corresponding to the second diagram is
\begin{align}
\frac{1}{\beta}\sum_{i q_n} \int \dbar \myvq \, \dbar \myvk \,
G_{\varphi}(\myvq+\myvk, i q_n) G_{\psi}(\myvq,i q_n)
\Delta^{\phantom{*}}_\mu(\myvk) \Delta^*_\nu(\myvk) (q+k/2)_{\mu}
(q+k/2)_{\nu}
\\
=\frac{1}{\beta}\sum_{i q_n} \int \dbar \myvq \, \dbar \myvk \,
\frac{ \Delta^{\phantom{*}}_\mu(\myvk) \Delta^*_\nu(\myvk)
(q+k/2)_{\mu} (q+k/2)_{\nu}} {(i q_n-(q^2-\mu-\jh))(i
q_n-(|\myvk+\myvq|^2-\mu+\jh)}.
\end{align}
On applying the standard Feynman trick and simplifying, the
previous amplitude becomes
\begin{align}
\frac{1}{\beta}\sum_{i q_n} \int \dbar \myvq \, \dbar \myvk \,
\int_0^1 d z \, \frac{ \Delta^{\phantom{*}}_{\mu}(\myvk)
\Delta^*_{\nu}(\myvk) (q+k/2)_{\mu} (q+k/2)_{\nu}}
{\left[(1-z)\left(i q_n-(q^2-\mu-\jh)\right)
      +z(i q_n-\left(|\myvk+\myvq|^2-\mu+\jh\right)\right]^2}
\\
\approx \frac{1}{\beta} \sum_{i q_n} \int \dbar \myvq \,\dbar
\myvk \,\int_0^1 dz \, \frac{ \Delta^{\phantom{*}}_\mu(\myvk)
\Delta^*_\nu(\myvk) q_\mu q_\nu} {\left[i q_n-q^2 + \mu - k^2 z +
k^2 z^2 + \jh(1-2z) \right]^2},
\end{align}
where in the final step we have dropped terms of higher order in
$\myvk$, as they do not contribute to the stiffness. This follows
from the obsevation that
$\Delta_{\alpha}^{*}\Delta_{\alpha}^{\phantom{*}}
=\partial_{\beta}\n \cdot \partial^{\beta}\n$. Reversing the order
of summation and integration, and simplifying further by noting
that the denominator sums to a $\delta$-function in the
zero-temperature limit, we obtain
\begin{align}
&&\int \dbar \myvq \, \dbar \myvk \, \int_0^1 dz
\,\Delta_\mu(\myvk) \Delta^*_\nu(\myvk) (q_\mu q_\nu) \delta\left[
-q^2 + \mu - k^2 z + k^2 z^2+\jh(1-2z)\right]
\\
&& =\frac{ -\Theta[\mu-\jh](\mu-\jh)^{1+d/2}
+\Theta[\jh+\mu](\jh+\mu)^{1+d/2}}{2^{2+d}
\pi^{d/2}\jh\,\Gamma[2+d/2]} \int d{\bf
x}\,\left(\partial_{\beta}\n\cdot\partial_{\beta}\n\right),
\end{align}
which simplifies to the expressions~(\ref{eq:smart}) in the text.
The other diagrams, which are more complicated are nevertheless
evaluated in a similar fashion, and since the technique is rather
well established and used commonly, we find it redundant to
present the details here.

\begin{figure}
\begin{tabular}{ccccc}
1.\includegraphics[width=2.2cm]{a1.eps}&
2.\includegraphics[width=2.8cm]{a_aa1.eps}&
3.\includegraphics[width=2.8cm]{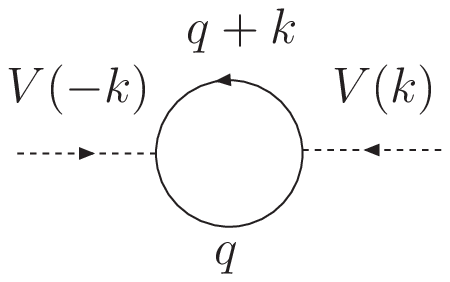}&
4.\includegraphics[width=2.8cm]{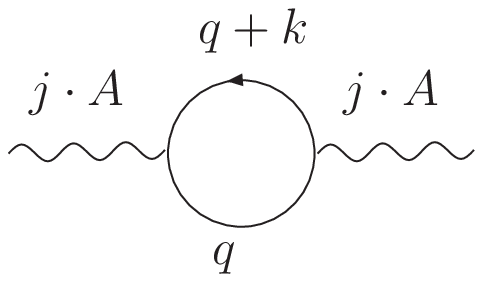}&
5.\includegraphics[width=2.8cm]{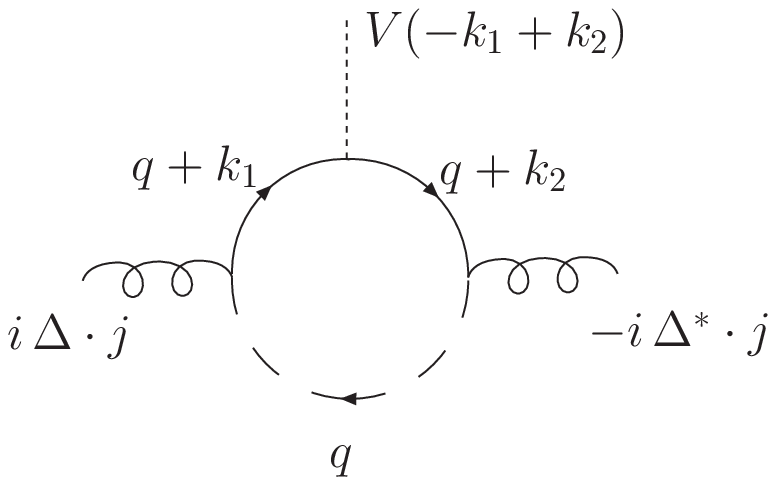}\\
6.\includegraphics[width=2.8cm]{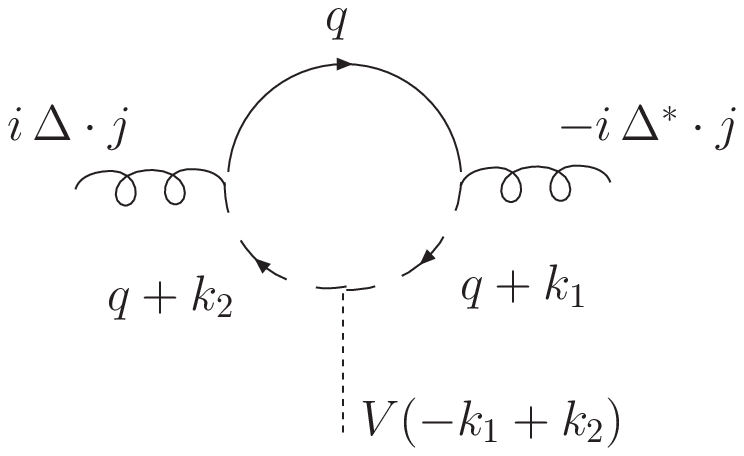}&
7.\includegraphics[width=2.8cm]{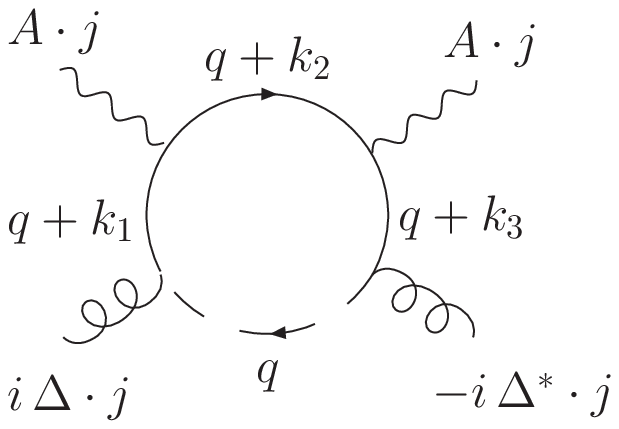}&
8.\includegraphics[width=2.8cm]{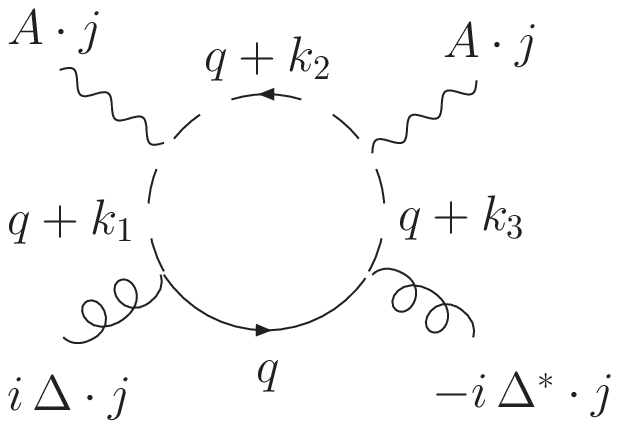}&
9.\includegraphics[width=2.8cm]{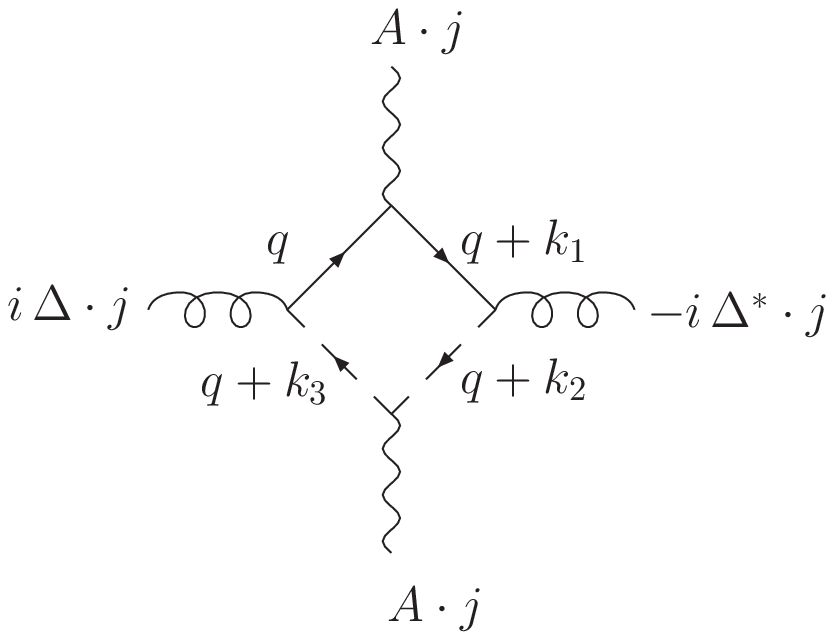}&
10.\includegraphics[width=2.5cm]{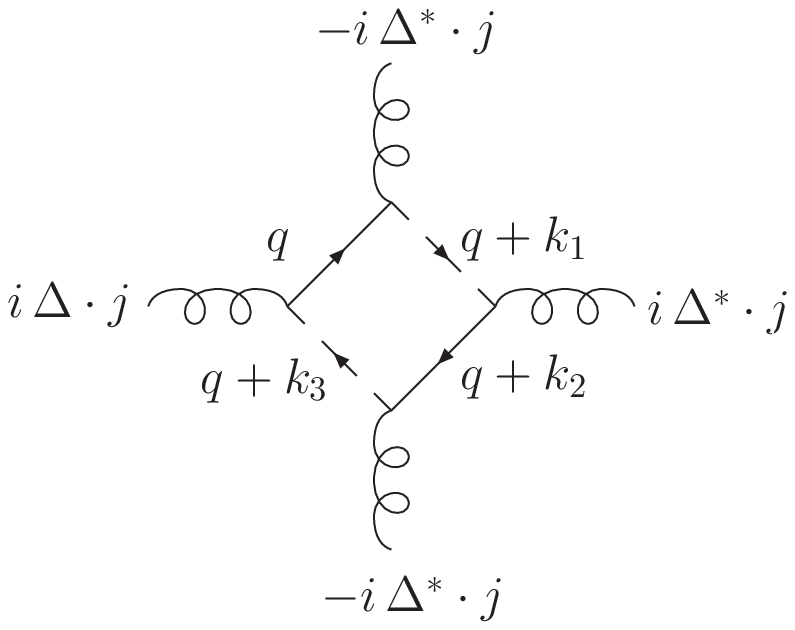}
\end{tabular}
\caption{Feynman diagrams contributing top the free energy, to fourth
order in gradients of the background spin texture.  See
Fig.~\ref{lowest-order} for the definitions of the propagators and
vertices.} \label{fig_diag}
\end{figure}
\end{widetext}

\end{document}